\documentclass[aps,prl,reprint,groupedaddress]{revtex4-1}

\usepackage{euscript,amsmath,amssymb,amsfonts,graphicx,bm,amsthm,mathtools}
  \usepackage{paralist}
  \usepackage{epstopdf}
  \usepackage{graphics} 
 \usepackage[latin1]{inputenc}
\usepackage[caption=false]{subfig}
\usepackage{soul}

\usepackage{latexsym,amssymb,enumerate,amsmath,amsthm} 
  \usepackage{graphicx} 
  \usepackage{tikz}
  \usepackage[hidelinks]{hyperref}
\usepackage{latexsym,amssymb,enumerate,amsmath} 
\usepackage{pgfplots}
\usepackage{soul,xcolor}

\newcommand{\grad}{\nabla}
\newcommand{\dd}{\textup{d}}
\def\eps{\varepsilon}
\def\E{\mathbb{E}}

\def\R{\mathbb{R}}
\theoremstyle{plain}
\theoremstyle{remark}

\theoremstyle{definition}

\begin{document}

%\preprint{}

\title[]{Escape from heterogeneous diffusion}


\author{Hwai-Ray Tung}
\affiliation{University of Utah, Department of Mathematics, Salt Lake City, UT 84112 USA}
\author{Sean D. Lawley}
\email[]{lawley@math.utah.edu}
\affiliation{University of Utah, Department of Mathematics, Salt Lake City, UT 84112 USA}


\date{\today}

\begin{abstract}
Many physical processes depend on the time it takes a diffusing particle to find a target. Though this classical quantity is now well-understood in various scenarios, little is known if the diffusivity depends on the location of the particle. For such heterogeneous diffusion, an ambiguity arises in interpreting the stochastic process, which reflects the well-known It\^{o} versus Stratonovich controversy. Here we analytically determine the mean escape time and splitting probabilities for an arbitrary heterogeneous diffusion in an arbitrary three-dimensional domain with small targets that can be perfectly or imperfectly absorbing. Our analysis reveals general principles for how search depends on heterogeneous diffusion and its interpretation (e.g.\ It\^{o}, Stratonovich, or kinetic). An intricate picture emerges in which, for instance, increasing the diffusivity can decrease, not affect, or even increase the escape time. Our results could be used to determine the appropriate interpretation for specific physical systems.
\end{abstract}

%\keywords{}

\maketitle

\textit{Introduction.}
How long does it take a diffusing particle to find a target? This classic ``search'' or ``escape'' question \cite{rayleigh1945, redner2001} continues to attract great interest \cite{bebon2023controlling, baravi2025first, biswas2025target}. Applications abound \cite{grebenkov2024target}, especially in chemistry \cite{hanggi1990} and biology \cite{bressloff2013stochastic, chou2014first}. 
Analyses typically focus on the mean first passage time (MFPT), which is the average time it takes the particle to be ``absorbed'' at a target. 
Many works study the MFPT when the targets are small windows on the boundary of a confining domain, which is the celebrated narrow escape problem \cite{benichou2008narrow, grebenkov2016, agranov2018narrow, holcman2014, ward10, ward10b}.

The paradigm of spatiotemporally constant diffusion has been extended to better model the complexities of natural and engineered systems. These extensions include a diffusivity which fluctuates stochastically in time \cite{chubynsky2014, sposini2024being, lawley2019dtmfpt}, intermittent switching between diffusion and other search modes \cite{benichou2005optimal, benichou2011rev, palyulin2016}, stochastic resetting \cite{reuveni2016optimal, pal2017first, de2020optimization, bressloff2020search}, as well as subdiffusive \cite{metzler1999, condamin2007first, grebenkov2010subdiffusion} and superdiffusive motion \cite{levernier2020, palyulin2019, gomez2024first}.

One natural and deceptively simple extension involves a diffusion coefficient which depends on the spatial location of the particle. 
For such ``heterogeneous diffusion'' \cite{vaccario2015} or ``nonlinear Brownian motion'' \cite{klimontovich1994nonlinear}, the position $X(t)$ of the particle at time $t$ follows a stochastic differential equation of the form
\begin{align}\label{eq:sde0}
    \dd X
    =\sqrt{2D(X)}\,\dd W,
\end{align}
where $D(x)$ is a position-dependent diffusivity and $\dd W$ is a Brownian increment. Owing to subtleties of stochastic integration, \eqref{eq:sde0} is merely a ``pre-equation'' \cite{van1981ito} until one specifies how to interpret the multiplicative noise. The interpretation is specified by choosing a parameter $\alpha\in[0,1]$ and evaluating the diffusivity in \eqref{eq:sde0} at the following weighted average of the current particle position $X(t)$ and its position at the next infinitesimal time $X(t+\dd t)$,
\begin{align}\label{eq:weighted}
    (1-\alpha)X(t)
    +\alpha X(t+\dd t).
\end{align}
The choice of $\alpha$ is reflected in the following forward Fokker-Planck equation governing the probability density $p(x,t)$ that $X(t)=x$,
\begin{align}\label{eq:fpe0}
    \partial_t p    =\grad\cdot[(D(x))^\alpha\grad[(D(x))^{1-\alpha}p]].
\end{align}
The most common choices are $\alpha=0$ (It\^{o} \cite{ito1944stochastic}), $\alpha=1/2$ (Stratonovich \cite{stratonovich1966new}), and $\alpha=1$ (kinetic or isothermal or H{\"a}nggi-Klimontovich \cite{hanggi1982stochastic}). Despite considerable debate and controversy regarding the It\^{o}-Stratonovich dilemma \cite{van1981ito, mannella2012ito, volpe2010influence, mannella2011comment, volpe2011volpe}, it is now generally agreed that there is no universally ``correct'' choice of $\alpha$ \cite{mannella2012ito}. Indeed, the It\^{o}, Stratonovich, and kinetic choices are each valid in different scenarios \cite{sokolov2010ito}. However, determining the appropriate choice of $\alpha$ for a given physical system is an area of current research \cite{pacheco2024langevin}. Furthermore,  except for some limited circumstances \cite{godec2015optimization, vaccario2015, godec2016first}, little is known about how $\alpha$ affects diffusive search.

%%%%%%%%%%%%%%%%%%%%%%%%%%%%%%%%
\begin{figure}[b]
\centering
\includegraphics[width=.7\linewidth]{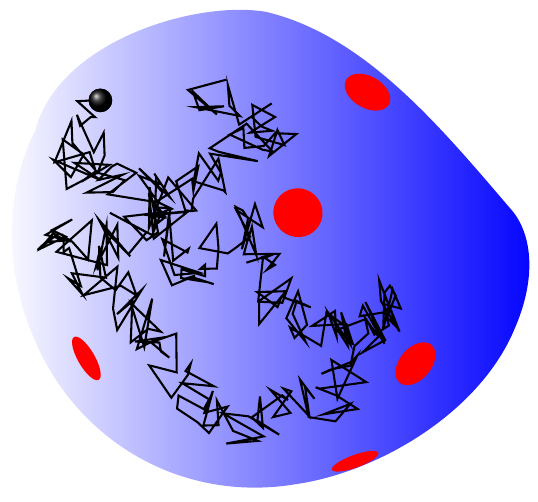}
\caption{A particle diffuses with a space-dependent diffusivity inside a general three-dimensional domain with small targets on its boundary.}
\label{fig:schem}
\end{figure}
%%%%%%%%%%%%%%%%%%%%%%%%%%%%%%%%

In this Letter, we study how heterogeneous diffusive search depends on $\alpha\in[0,1]$. We consider a three-dimensional diffusion process $X$ following \eqref{eq:sde0}-\eqref{eq:fpe0} with an arbitrary space-dependent diffusivity $D(x)$ in an arbitrary bounded domain containing small targets on its otherwise reflecting boundary (illustrated in Fig.~\ref{fig:schem}). We derive analytical formulas for the MFPT and the probability that the particle reaches a given target (the so-called splitting probability \cite{redner2001}). We consider both perfect targets (absorption occurs immediately upon first encounter) and imperfect targets (absorption occurs with some probability upon each encounter \cite{grebenkov2020paradigm, guerin2023imperfect}). Our results reveal that the MFPT and splitting probability can depend strongly and counterintuitively on the multiplicative noise interpretation.

%%%%%%%%%%%%%%%%%%%%%%%%%%%%%%%%
\textit{Mathematical analysis.}
Let the heterogeneous diffusion $X$ described by \eqref{eq:sde0}-\eqref{eq:fpe0} diffuse in a bounded domain $\Omega\subset\R^3$. Suppose the smooth boundary $\partial\Omega$ is reflecting except for $N\ge1$ well-separated disks $\partial\Omega_1,\dots,\partial\Omega_N$ of radius $a>0$, and thus the density $p$ in \eqref{eq:fpe0} satisfies the following mixed boundary conditions,
\begin{align}
    D(x)^\alpha\grad[D(x)^{1-\alpha}p]\cdot\mathbf{n}
    &=0,\quad x\in\partial\Omega\backslash\{\cup_{n=1}^N\partial\Omega_n\},\label{eq:preflect0}\\
    p
    &=0,\quad x\in\cup_{n=1}^N\partial\Omega_n,\label{eq:pabsorb0}
\end{align}
where $\mathbf{n}=\mathbf{n}(x)$ is the unit normal at $x\in\partial\Omega$. Solving \eqref{eq:fpe0}-\eqref{eq:pabsorb0} with initial position $X(0)=x_0$ in the open set $\Omega$ via a spectral expansion yields
\begin{align}\label{eq:spectral}
    p(x,t)
    &=\frac{D(x_0)^{1-\alpha}}{\int_\Omega D(y)^{\alpha-1}\,\dd y}\sum_{k\ge0}e^{-\lambda_k t}u_k(x_0) {u}_k(x),
\end{align}
where $0<\lambda_1<\cdots$ are the eigenvalues of the negative of the spatial differential operator in \eqref{eq:fpe0},
\begin{align}\label{eq:feval0}
    \lambda_k u_k
    =-\grad\cdot[(D(x))^\alpha\grad[(D(x))^{1-\alpha}u_k]],\quad x\in\Omega,
\end{align}
and the eigenfunctions $u_k$ satisfy the boundary conditions in \eqref{eq:preflect0}-\eqref{eq:pabsorb0} and are orthogonal under the following weighted inner product,
\begin{align*}
    (f,g)
    :=\int_\Omega f(x)g(x)D(x)^{1-\alpha}\,\dd x,
\end{align*}
and are normalized so that $(u_k,u_k)=\int_\Omega D(x)^{\alpha-1}\,\dd x$. 

If $\eps:=a/|\Omega|^{1/3}$ compares the target radius to the domain length scale, then the principal eigenvalue vanishes and the principal eigenfunction approaches the stationary density of the targetless domain in the small target limit,
\begin{align}\label{eq:key}
    \lambda_0\to0\;\text{ and }\;u_0(x)\to D(x)^{\alpha-1}\quad\text{as }\eps\to0.
\end{align}
It follows immediately from \eqref{eq:spectral} and \eqref{eq:key} that the MFPT diverges as the reciprocal of the principal eigenvalue,
\begin{align}\label{eq:Teval}
    T&:=\int_0^\infty \int_\Omega p(x,t)\,\dd x\,\dd t
    \sim 1/\lambda_0\quad\text{as }\eps\to0.
\end{align}
The approximation $T\approx1/\lambda_0$ requires $\eps\ll1$ and the initial location $X(0)=x_0$ to be outside an $\mathcal{O}(a)$ neighborhood of any target ($X(0)$ could also be uniformly distributed in $\Omega$ since then the probability that $X(0)$ is near a target is $\mathcal{O}(\eps)$ as $\eps\to0$). 

Integrating \eqref{eq:feval0} over $\Omega$ for $k=0$ and using the divergence theorem, the boundary condition \eqref{eq:preflect0}, and the relation \eqref{eq:Teval} implies that the MFPT is given by
\begin{align}\label{eq:Tfromflux}
    T
    \sim\frac{\int_\Omega D(x)^{\alpha-1}\,\dd x}{-\sum_{n=1}^N\int_{\partial\Omega_n}D(x)^\alpha\partial_{\mathbf{n}}[D(x)^{1-\alpha}u_0]\,\dd x}\quad\text{as }\eps\to0.
\end{align}
If the diffusivity $D(x)$ is continuous in a neighborhood of each target, then strong localized perturbation theory \cite{ward1993} yields the flux into a target,
\begin{align}\label{eq:flux}
    -\int_{\partial\Omega_n}D(x)^\alpha\partial_{\mathbf{n}}[D(x)^{1-\alpha}u_0]\,\dd x
    \sim4aD(x_n)^\alpha\quad\text{as }\eps\to0,
\end{align}
where $x_n\in\partial\Omega$ is the center of the $n$th target. 

\begin{figure}%[b]
\centering
\includegraphics[width=1\linewidth]{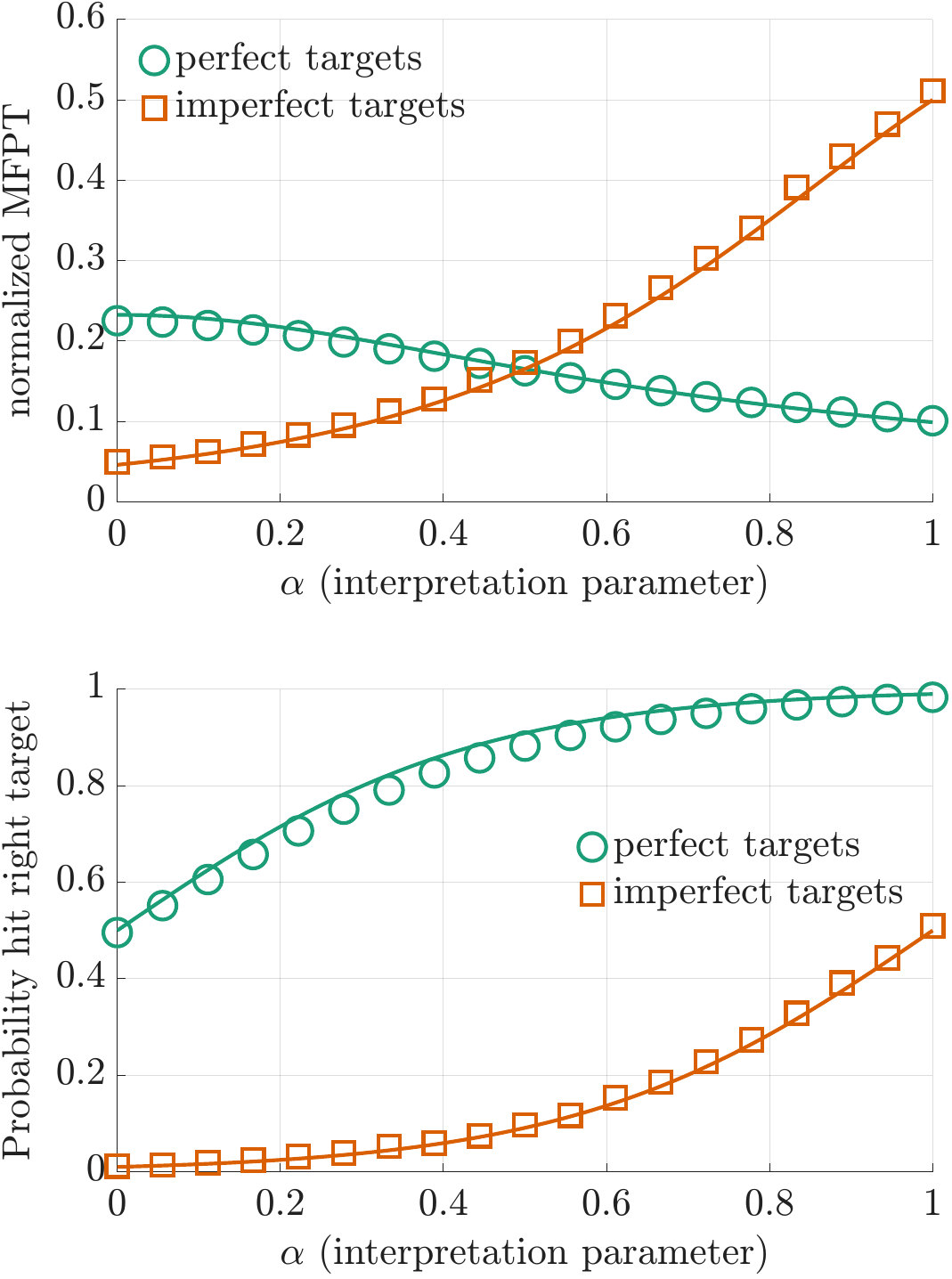}
\caption{Comparison between theory (curves) and stochastic simulations (markers). Simulated particles start in the center of a unit cube domain $\Omega$ with $N=2$ targets with common radius $a=0.01$ located at the center of the left and right boundaries. The diffusivity is $D(x)=D(x^{(1)},x^{(2)},x^{(3)})=0.1+10x^{(1)}$, where $x^{(1)}=0$ and $x^{(1)}=1$ define the left and right boundaries. The MFPTs in the top panel are multiplied by $4a$ for perfect targets and $\kappa\pi a^2$ for imperfect targets. The imperfect targets have reactivity $\kappa=1$ \cite{data_availability}.}
\label{fig:sim}
\end{figure}

%%%%%%%%%%%%%%%%%%%%%%%%%%%%%%%%%%%%%%%%%%%%%%%%%%%%%%%%%%%%%%%%%%%%%%%%%%%
\textit{MFPT to perfect targets.} 
Combining \eqref{eq:Tfromflux} and \eqref{eq:flux} yields the MFPT,
\begin{align}\label{eq:main1}
    T
    \sim\frac{1}{4a}\frac{\int_\Omega D(x)^{\alpha-1}\,\dd x}{\sum_{n=1}^N D(x_n)^\alpha}\quad\text{as }\eps\to0,
\end{align}
which is our first main result. Fig.~\ref{fig:sim} shows close agreement between the theory in \eqref{eq:main1} and stochastic simulations. 

Notice that if $\alpha=0$ (It\^{o}), then the MFPT depends on the diffusivity only via the globally averaged reciprocal diffusivity,
\begin{align}\label{eq:Tito0}
    T_{\text{ito}}
    =\frac{1}{4aN}\int_\Omega D(x)^{-1}\,\dd x.
\end{align}
Furthermore, as $\alpha$ increases, \eqref{eq:main1} implies that the globally averaged diffusivity becomes less important, and instead the diffusivity at each target more strongly influences the MFPT. In fact, if $\alpha=1$ (kinetic), then the MFPT depends only on the diffusivity at the targets,
\begin{align}\label{eq:Tkin0}
    T_{\text{kin}}
    =\frac{1}{4a}\frac{|\Omega|}{\sum_{n=1}^N D(x_n)}.
\end{align}
Hence, the MFPT for a kinetic interpretation is unaffected by changes to the diffusivity away from the targets. 
Notice also that if $\alpha>0$, then the MFPT is minimized if the targets are in regions of high diffusivity.

%%%%%%%%%%%%%%%%%%%%%%%%%%%%%%%%
%%%%%%%%%%%%%%%%%%%%%%%%%%%%%%%%

%%%%%%%%%%%%%%%%%%%%%%%%%%%%%%%%%%%%%%%%%%%%%%%%%%%%%%%%%%%%%%%%%%%%%%%%%%%
\textit{MFPT to imperfect targets.} 
A very different picture emerges if the targets are ``imperfect'' \cite{collins1949diffusion, grebenkov2020paradigm}, which means that the diffusing particle may not be absorbed upon its first encounter with a target. Such imperfect or ``partially absorbing'' targets model a variety of microscopic scenarios \cite{guerin2023imperfect}, including if absorption only occurs in certain particle orientations \cite{berg1985orientation, plunkett2021bimolecular}, a heterogeneous target (or particle \cite{lawley2019boundary}) containing both absorbing and reflecting regions \cite{berg1977physics}, a stochastically gated target that switches in time between absorbing (open) and reflecting (closed) \cite{benichou2000kinetics, lawley2015new}, and energetic \cite{shoup1982role} or entropic \cite{zhou1991rate} activation barriers. 
Precisely, imperfect targets are defined by replacing the absorbing boundary condition in \eqref{eq:pabsorb0} by the following partially absorbing condition,
\begin{align}\label{eq:ppartial}
    -D(x)^\alpha\grad[D(x)^{1-\alpha}p]\cdot\mathbf{n}
    =\kappa p,\quad x\in\cup_{n=1}^N\partial\Omega_n,
\end{align}
where the parameter $\kappa>0$ measures target reactivity.

It follows immediately from \eqref{eq:Tfromflux} that the MFPT is now
\begin{align*}
    T^{\text{imperf}}
    \sim\frac{1}{\kappa\pi a^2}\frac{\int_\Omega (D(x))^{\alpha-1}\,\dd x}{\sum_{n=1}^N(D(x_n))^{\alpha-1}}\quad\text{as }\eps\to0,
\end{align*}
which is our second main result.
The MFPT for imperfect targets $T^{\text{imperf}}$ differs sharply from the MFPT for perfect targets $T$ in four salient ways. First, the diffusivity at the targets $D(x_n)$ influences the MFPT for all $\alpha\in[0,1)$, and this influence is greater for smaller $\alpha$. Indeed, the value of the diffusivity near targets has the strongest influence if the multiplicative noise is It\^{o},
\begin{align}\label{eq:Titoimperf0}
    T_{\text{ito}}^{\text{imperf}}
    =\frac{1}{\kappa\pi a^2}\frac{\int_\Omega 1/D(x)\,\dd x}{\sum_{n=1}^N 1/D(x_n)}.
\end{align}
Second, increasing the diffusivity at the targets increases the MFPT if $\alpha\in[0,1)$. Third, the MFPT is minimized if the targets are in regions of low diffusivity. Fourth, the MFPT is independent of the diffusivity if $\alpha=1$,
\begin{align}\label{eq:Tkinimperf0}
    T_{\text{kin}}^{\text{imperf}}
    =\frac{|\Omega|}{\kappa\pi a^2N}.
\end{align}

%%%%%%%%%%%%%%%%%%%%%%%%%%%%%%%%%%%%%%%%%%%%%%%%%%%%%%%%%%%%%%%%%%%%%%%%%%%
\textit{Splitting probability.} 
In addition to the MFPT, another important statistic of diffusive escape is the so-called splitting probability \cite{redner2001}, which is the probability that the particle escapes via a particular target. The probability of exit through the $j$th target is obtained by integrating the flux of the density $p$ through that target,
\begin{align}
    P&:=-\int_0^\infty\int_{\partial\Omega_j}D^\alpha\grad[D^{1-\alpha}p]\cdot \mathbf{n}\,\dd x\,\dd t\nonumber\\
    &\sim\frac{\int_{\partial\Omega_j}D^\alpha\grad[D^{1-\alpha}u_0]\cdot\mathbf{n}\,\dd x}{\sum_{n=1}^N\int_{\partial\Omega_n}D^\alpha\grad[D^{1-\alpha}u_0]\cdot\mathbf{n}\,\dd x}\quad\text{as }\eps\to0,\label{eq:Pasymptotic}
\end{align}
where the small target asymptotic in \eqref{eq:Pasymptotic} follows from \eqref{eq:spectral} and \eqref{eq:key}. 

If the targets are perfect, then \eqref{eq:flux} and \eqref{eq:Pasymptotic} yield the following splitting probability,
\begin{align}\label{eq:main3}
    P
    \to\frac{D(x_j)^\alpha}{\sum_{n=1}^N D(x_n)^\alpha}\quad\text{as }\eps\to0,
\end{align}
which is our third main result. Notice first that $P$ is uniform on the targets if $\alpha=0$,
\begin{align}\label{eq:Pito0}
    P_{\text{ito}}
    =1/N.
\end{align}
Second, if $\alpha\neq0$, then only the diffusivity at the targets $D(x_n)$ influences $P$, and this influence increases as $\alpha$ increases. Third, targets in regions of high diffusivity are more likely to be hit by the particle.

The opposite three conclusions hold for imperfect targets. Indeed, \eqref{eq:ppartial} and \eqref{eq:Pasymptotic} imply that the splitting probability for imperfect targets is
\begin{align*}
    P^{\text{imperf}}
    \to\frac{D(x_j)^{\alpha-1}}{\sum_{n=1}^N D(x_n)^{\alpha-1}}\quad\text{as }\eps\to0.
\end{align*}
Hence, $P^{\text{imperf}}$ is uniform on the targets if $\alpha=1$,
\begin{align}\label{eq:Pkinimperf0}
    P_{\text{kin}}^{\text{imperf}}
    =1/N,
\end{align}
which is identical to the It\^{o} case for perfect targets in \eqref{eq:Pito0}. 
Further, if $\alpha\neq1$, then only the diffusivity at the targets $D(x_n)$ influences $P$, and this influence decreases as $\alpha$ increases. Finally, targets in regions of high diffusivity are less likely to be hit by the particle.

%%%%%%%%%%%%%%%%%%%%%%%%%%%%%%%%%%%%%%%%%%%%%%%%%%%%%%%%%%%%%%%%%%%%%%%%%%%%%%%%%%%%%%%%%%%%%%%%%%%%%%%%%%%%%%%%%%%%%%%%%%%%%
\textit{Rationalizing the results.} 
We now use stochastic calculus to rationalize our results.

If $\alpha=0$ (It\^{o}), then $X$ is a time-changed Brownian motion. That is, the particle is unbiased and simply slows down in regions of low diffusivity. Hence, the MFPT $T_{\text{ito}}$ in \eqref{eq:Tito0} is proportional to the time scale of domain exploration and is agnostic to whether the targets are located in regions of high or low diffusivity. Furthermore, particle speed does not affect which target it ultimately finds, and thus $P_{\text{ito}}$ in \eqref{eq:Pito0} is independent of the diffusivity and is uniform on the $N$ identically-shaped targets.

If $\alpha>0$, then the drift-free stochastic differential equation in \eqref{eq:sde0} can be converted to the following It\^{o} form with a drift that ``pushes up'' the diffusivity gradient,
\begin{align}\label{eq:convert}
    \dd X
    =\alpha \grad D(X)\,\dd t+\sqrt{2D(X)}\cdot\dd W,
\end{align}
where ``$\sqrt{2D(X)}\cdot\dd W$'' denotes multiplicative noise with It\^{o} interpretation. Hence, the particle is biased to move toward regions of high diffusivity. This is why the MFPT in \eqref{eq:main1} depends more strongly on the diffusivity near the targets if $\alpha>0$ and why the splitting probability in \eqref{eq:main3} is larger for targets in regions of high diffusivity. Nevertheless, it is perhaps still unexpected that the $\alpha=1$ MFPT $T_{\text{kin}}$ in \eqref{eq:Tkin0} and the $\alpha\neq0$ splitting probability in \eqref{eq:main3} are both independent of the diffusivity away from the targets.

Our results for imperfect targets are roughly opposite of our results for perfect targets. To understand this discrepancy, recall that absorption at an imperfect target requires the particle to be near the target for a sufficient amount of time (precisely, the boundary local time on the target must exceed an independent exponentially distributed random variable \cite{grebenkov2020paradigm}). Thus, increasing the diffusivity near a target has the two opposing effects of (i) accelerating particle motion, which decreases the time near the target and thus decreases the likelihood of absorption, and (ii) biasing the particle to move toward the target if $\alpha>0$, which increases the likelihood of absorption.

If $\alpha=0$, then only effect (i) is relevant, which explains why increasing the diffusivity near targets increases the It\^{o} MFPT $T_{\text{ito}}^{\text{imperf}}$ in \eqref{eq:Titoimperf0} and why targets in regions of high diffusivity are less likely to be found. As $\alpha$ increases away from 0, effect (ii) begins to cancel effect (i), which explains why the diffusivity near targets has a weaker influence on the MFPT and splitting probability for larger $\alpha$. If $\alpha=1$, then effects (i) and (ii) perfectly cancel, which is why the kinetic MFPT $T_{\text{kin}}^{\text{imperf}}$ in \eqref{eq:Tkinimperf0} and the kinetic splitting probability $P_{\text{kin}}^{\text{imperf}}$ in \eqref{eq:Pkinimperf0} are both independent of the diffusivity.

%%%%%%%%%%%%%%%%%%%%%%%%%%%%%%%%%%%%%%%%%%%%
\textit{Discussion.} 
Despite decades of interest, debate, and controversy regarding heterogeneous diffusion \cite{van1981ito, mannella2012ito, volpe2010influence, mannella2011comment, volpe2011volpe, sokolov2010ito}, the effects of such heterogeneity on diffusive escape have largely been unknown. Important advances in this area were made by Grebenkov \cite{grebenkov2016} for the It\^{o} interpretation in a two-dimensional domain with one perfect target, as well as by Godec and Metzler \cite{godec2015optimization, godec2016first} and Vaccario, Antoine, and Talbot \cite{vaccario2015} for quasi-one-dimensional domains with a single target and a simple diffusivity profile. Specifically, Refs.~\cite{godec2015optimization, vaccario2015} derived exact results on the MFPT (and mean residence time, see below) for a single perfect target in the center of a radially symmetric domain with a radially symmetric and piecewise constant diffusivity $D(x)$. In contrast, we derived general asymptotic results on the MFPT and splitting probability in an arbitrary domain with multiple small perfect or imperfect targets and an arbitrary diffusivity $D(x)$.

Our analysis elucidates general principles for how the multiplicative noise parameter $\alpha\in[0,1]$ affects escape statistics. Our results could thus be used to determine the appropriate value of $\alpha$ for a particular physical system. 
Indeed, while an important recent study by Pacheco-Pozo et al.~\cite{pacheco2024langevin} proposed a method to assign a value of $\alpha$ to a one-dimensional experimental system if the diffusivity is piecewise constant with only two possible values (a so-called two-phase system), our analysis provides a basis for methods and guidelines to choose $\alpha$ in more general scenarios.

Our approach in this Letter can be extended to obtain more detailed statistics about even more general diffusion processes. For instance, \eqref{eq:spectral} and \eqref{eq:key}-\eqref{eq:Teval} imply that our results on the MFPT $T$ are readily extended to the mean residence time $T_{\Omega'}$ in any subdomain $\Omega'\subset\Omega$. Indeed, if the targets are perfect, then
\begin{align*}
    T_{\Omega'}
    :=\int_0^\infty\int_{\Omega'}p\,\dd x\,\dd t
    \sim\frac{1}{4a}\frac{\int_{\Omega'} D(x)^{\alpha-1}\,\dd x}{\sum_{n=1}^N D(x_n)^\alpha}\quad\text{as }\eps\to0,
\end{align*}
and if the targets are imperfect, then
\begin{align*}
    T_{\Omega'}
    \sim\frac{1}{\kappa\pi a^2}\frac{\int_{\Omega'} (D(x))^{\alpha-1}\,\dd x}{\sum_{n=1}^N(D(x_n))^{\alpha-1}}\quad\text{as }\eps\to0.
\end{align*}
In addition, the $m$th moment of the stochastic FPT $\tau$ is given by $\E[\tau^m]\sim(m!)T^m$ and $\tau/T$ converges to a unit mean exponential random variable as $\eps\to0$. Furthermore, our results can be extended to targets of arbitrary shapes, targets in the interior of the domain, two-dimensional domains, targets which are not necessarily small but instead have low reactivity (in any spatial dimension), and higher order asymptotic corrections. 
These extensions will be the subject of a forthcoming full-length article.

%%%%%%%%%%%%%%%%%%%%%%%%%%%%%%%%%%%%%%%%%%%%%%%%%%%%%%%%%%%%%%%%%%%%%%%%%%%%%%%%%%%%%%%%%%%%%%%%%%%%%%%%%%%%%%%%%%%%%%%%%%%%%%%%%%%%%%%%%%%%
\subsubsection*{Acknowledgments}
SDL was supported by the National Science Foundation (Grant Nos.\ CAREER DMS-1944574 and DMS-2325258).

%%%%%%%%%%%%%%%%%%%%%%%%%%%%%%%%%%%%%%%%%%%%%%%%%%%%%%%%%%%%%%%%%%%%%%%%%%%%%%%%%%%%%%%%%%%%%%%%%%%%%%%%%%%%%%%%%%%%%%%%%%%%%%%%%%%%%%%%%%%%

\bibliography{library.bib}

@misc{data_availability,
  note = {Simulation code is available at \url{https://github.com/seanlawley/Dx}.}
}

@article{bressloff2020search,
  title={Search processes with stochastic resetting and multiple targets},
  author={Bressloff, Paul C},
  journal={Physical Review E},
  volume={102},
  number={2},
  pages={022115},
  year={2020},
  publisher={APS}
}

@article{mannella2011comment,
  title={Comment on ``Influence of Noise on Force Measurements''},
  author={Mannella, Riccardo and McClintock, PVE},
  journal={Physical Review Letters},
  volume={107},
  number={7},
  pages={078901},
  year={2011},
  publisher={APS}
}

@article{volpe2010influence,
  title={Influence of noise on force measurements},
  author={Volpe, Giovanni and Helden, Laurent and Brettschneider, Thomas and Wehr, Jan and Bechinger, Clemens},
  journal={Physical Review Letters},
  volume={104},
  number={17},
  pages={170602},
  year={2010},
  publisher={APS}
}

@article{volpe2011volpe,
  title={Volpe et al. reply},
  author={Volpe, Giovanni and Helden, Laurent and Brettschneider, Thomas and Wehr, Jan and Bechinger, Clemens},
  journal={Physical Review Letters},
  volume={107},
  number={7},
  pages={078902},
  year={2011},
  publisher={APS}
}

@article{zhou1991rate,
  title={A rate process with an entropy barrier},
  author={Zhou, Huan-Xiang and Zwanzig, Robert},
  journal={The Journal of chemical physics},
  volume={94},
  number={9},
  pages={6147--6152},
  year={1991},
  publisher={American Institute of Physics}
}

@article{benichou2000kinetics,
  title = {Kinetics of stochastically gated diffusion-limited reactions and geometry of random walk trajectories},
  author = {B\'enichou, O. and Moreau, M. and Oshanin, G.},
  journal = {Phys. Rev. E},
  volume = {61},
  issue = {4},
  pages = {3388--3406},
  numpages = {0},
  year = {2000},
  month = {Apr},
  publisher = {American Physical Society},
  doi = {10.1103/PhysRevE.61.3388},
  url = {https://link.aps.org/doi/10.1103/PhysRevE.61.3388}
}

@article{plunkett2021bimolecular,
  title={Bimolecular binding rates for pairs of spherical molecules with small binding sites},
  author={Plunkett, Claire E and Lawley, Sean D},
  journal={Multiscale Modeling \& Simulation},
  volume={19},
  number={1},
  pages={148--183},
  year={2021},
  publisher={SIAM}
}

@book{grebenkov2024target,
  title={Target search problems},
  author={Grebenkov, Denis and Metzler, Ralf and Oshanin, Gleb},
  pages={1--29},
  year={2024},
  publisher={Springer}
}

@article{benichou2005optimal,
  title={Optimal search strategies for hidden targets},
  author={B{\'e}nichou, O and Coppey, M and Moreau, M and Suet, PH and Voituriez, R},
  journal={Physical Review Letters},
  volume={94},
  number={19},
  pages={198101},
  year={2005},
  publisher={APS}
}

@article{reuveni2016optimal,
  title={Optimal stochastic restart renders fluctuations in first passage times universal},
  author={Reuveni, Shlomi},
  journal={Physical Review Letters},
  volume={116},
  number={17},
  pages={170601},
  year={2016},
  publisher={APS}
}

@article{de2020optimization,
  title={Optimization in first-passage resetting},
  author={De Bruyne, B and Randon-Furling, Julien and Redner, S},
  journal={Physical Review Letters},
  volume={125},
  number={5},
  pages={050602},
  year={2020},
  publisher={APS}
}

@article{biswas2025target,
title = {Target Search Optimization by Threshold Resetting},
  author = {Biswas, Arup and Majumdar, Satya N. and Pal, Arnab},
  journal = {Physical Review Letters},
  volume = {135},
  issue = {22},
  pages = {227101},
  numpages = {8},
  year = {2025},
  month = {Nov},
  publisher = {American Physical Society},
  doi = {10.1103/752c-wqly},
  url = {https://link.aps.org/doi/10.1103/752c-wqly}
}

@article{pal2017first,
  title={First passage under restart},
  author={Pal, Arnab and Reuveni, Shlomi},
  journal={Physical Review Letters},
  volume={118},
  number={3},
  pages={030603},
  year={2017},
  publisher={APS}
}

@article{baravi2025first,
  title={First passage times in compact domains exhibit biscaling},
  author={Baravi, Talia and Kessler, David A and Barkai, Eli},
  journal={Physical Review Letters},
  volume={134},
  number={12},
  pages={127101},
  year={2025},
  publisher={APS}
}

@article{bebon2023controlling,
  title={Controlling uncertainty of empirical first-passage times in the small-sample regime},
  author={Bebon, Rick and Godec, Alja{\v{z}}},
  journal={Physical Review Letters},
  volume={131},
  number={23},
  pages={237101},
  year={2023},
  publisher={APS}
}

@article{agranov2018narrow,
  title={Narrow escape of interacting diffusing particles},
  author={Agranov, Tal and Meerson, Baruch},
  journal={Physical Review Letters},
  volume={120},
  number={12},
  pages={120601},
  year={2018},
  publisher={APS}
}

@article{godec2016first,
  title={First passage time distribution in heterogeneity controlled kinetics: going beyond the mean first passage time},
  author={Godec, Alja{\v{z}} and Metzler, Ralf},
  journal={Scientific reports},
  volume={6},
  number={1},
  pages={20349},
  year={2016},
  publisher={Nature Publishing Group UK London}
}

@article{pacheco2024langevin,
  title={Langevin equation in heterogeneous landscapes: how to choose the interpretation},
  author={Pacheco-Pozo, Adrian and Balcerek, Micha{\l} and Wy{\l}omanska, Agnieszka and Burnecki, Krzysztof and Sokolov, Igor M and Krapf, Diego},
  journal={Physical Review Letters},
  volume={133},
  number={6},
  pages={067102},
  year={2024},
  publisher={APS}
}

@article{condamin2007first,
  title={First-passage time distributions for subdiffusion in confined geometry},
  author={Condamin, S and B{\'e}nichou, O and Klafter, J},
  journal={Physical Review Letters},
  volume={98},
  number={25},
  pages={250602},
  year={2007},
  publisher={APS}
}

@article{sposini2024being,
  title={Being heterogeneous is advantageous: Extreme Brownian non-Gaussian searches},
  author={Sposini, Vittoria and Nampoothiri, Sankaran and Chechkin, Aleksei and Orlandini, Enzo and Seno, Flavio and Baldovin, Fulvio},
  journal={Physical Review Letters},
  volume={132},
  number={11},
  pages={117101},
  year={2024},
  publisher={APS}
}

@article{sokolov2010ito,
  title={{Ito, Stratonovich, H{\"a}nggi and all the rest: The thermodynamics of interpretation}},
  author={Sokolov, Igor M},
  journal={Chemical Physics},
  volume={375},
  number={2-3},
  pages={359--363},
  year={2010},
  publisher={Elsevier}
}

@article{klimontovich1994nonlinear,
  title={Nonlinear brownian motion},
  author={Klimontovich, Yu L},
  journal={Physics-Uspekhi},
  volume={37},
  number={8},
  pages={737},
  year={1994},
  publisher={IOP Publishing}
}

@article{godec2015optimization,
  title={Optimization and universality of Brownian search in a basic model of quenched heterogeneous media},
  author={Godec, Alja{\v{z}} and Metzler, Ralf},
  journal={Physical Review E},
  volume={91},
  number={5},
  pages={052134},
  year={2015},
  publisher={APS}
}

@article{hanggi1982stochastic,
  title={Stochastic processes: Time evolution, symmetries and linear response},
  author={H{\"a}nggi, Peter and Thomas, Harry},
  journal={Physics Reports},
  volume={88},
  number={4},
  pages={207--319},
  year={1982},
  publisher={Elsevier}
}

@article{benichou2008narrow,
  title={Narrow-escape time problem: Time needed for a particle to exit a confining domain through a small window},
  author={B{\'e}nichou, O and Voituriez, R},
  journal={Physical Review Letters},
  volume={100},
  number={16},
  pages={168105},
  year={2008},
  publisher={APS}
}

@article{guerin2023imperfect,
  title={Imperfect narrow escape problem},
  author={Gu{\'e}rin, Thomas and Dolgushev, M and B{\'e}nichou, O and Voituriez, R},
  journal={Physical Review E},
  volume={107},
  number={3},
  pages={034134},
  year={2023},
  publisher={APS}
}

@article{stratonovich1966new,
  title={A new representation for stochastic integrals and equations},
  author={Stratonovich, RL},
  journal={SIAM Journal on Control},
  volume={4},
  number={2},
  pages={362--371},
  year={1966},
  publisher={SIAM}
}

@article{ito1944stochastic,
  title={Stochastic integral},
  author={It{\^o}, Kiyosi},
  journal={Proceedings of the Imperial Academy},
  volume={20},
  number={8},
  pages={519--524},
  year={1944},
  publisher={The Japan Academy}
}

@article{mannella2012ito,
  title={It{\^o} versus Stratonovich: 30 years later},
  author={Mannella, Riccardo and McClintock, Peter VE},
  journal={Fluctuation and Noise Letters},
  volume={11},
  number={01},
  pages={1240010},
  year={2012},
  publisher={World Scientific}
}

@article{van1981ito,
  title={It{\^o} versus stratonovich},
  author={Van Kampen, Nicolaas G},
  journal={Journal of Statistical Physics},
  volume={24},
  number={1},
  pages={175--187},
  year={1981},
  publisher={Springer}
}

@article{grebenkov2020paradigm,
  title={Paradigm shift in diffusion-mediated surface phenomena},
  author={Grebenkov, Denis S},
  journal={Physical Review Letters},
  volume={125},
  number={7},
  pages={078102},
  year={2020},
  publisher={APS}
}

@article{palyulin2016,
  title={{Search reliability and search efficiency of combined L{\'e}vy--Brownian motion: long relocations mingled with thorough local exploration}},
  author={Palyulin, Vladimir V and Chechkin, Aleksei V and Klages, Rainer and Metzler, Ralf},
  journal={Journal of Physics A: Mathematical and Theoretical},
  volume={49},
  number={39},
  pages={394002},
  year={2016},
  publisher={IOP Publishing}
}

@article{gomez2024first,
  title={First Hitting Time of a One-Dimensional L{\'e}vy Flight to Small Targets},
  author={Gomez, Daniel and Lawley, Sean D},
  journal={SIAM Journal on Applied Mathematics},
  volume={84},
  number={3},
  pages={1140--1162},
  year={2024},
  publisher={SIAM}
}

@article{metzler1999,
  title={Anomalous diffusion and relaxation close to thermal equilibrium: A fractional {Fokker-Planck} equation approach},
  author={Metzler, Ralf and Barkai, Eli and Klafter, Joseph},
  journal={Physical Review Letters},
  volume={82},
  number={18},
  pages={3563},
  year={1999},
  publisher={APS}
}

@article{grebenkov2010subdiffusion,
  title={Subdiffusion in a bounded domain with a partially absorbing-reflecting boundary},
  author={Grebenkov, Denis S},
  journal={Physical review E},
  volume={81},
  number={2},
  pages={021128},
  year={2010},
  publisher={APS}
}

@article{lawley2019boundary,
  title={Boundary homogenization for trapping patchy particles},
  author={Lawley, Sean D},
  journal={Physical Review E},
  volume={100},
  number={3},
  pages={032601},
  year={2019},
  publisher={APS}
}

@article{hanggi1990,
  title={Reaction-rate theory: fifty years after Kramers},
  author={H{\"a}nggi, Peter and Talkner, Peter and Borkovec, Michal},
  journal={Reviews of modern physics},
  volume={62},
  number={2},
  pages={251},
  year={1990},
  publisher={APS}
}

@article{palyulin2019,
  title={First passage and first hitting times of {L}{\'e}vy flights and {L}{\'e}vy walks},
  author={Palyulin, Vladimir V and Blackburn, George and Lomholt, Michael A and Watkins, Nicholas W and Metzler, Ralf and Klages, Rainer and Chechkin, Aleksei V},
  journal={New Journal of Physics},
  volume={21},
  number={10},
  pages={103028},
  year={2019},
  publisher={IOP Publishing}
}

@article{levernier2020,
  title={Inverse square {L}{\'e}vy walks are not optimal search strategies for $d\ge 2$},
  author={Levernier, Nicolas and Textor, Johannes and B{\'e}nichou, Olivier and Voituriez, Rapha{\"e}l},
  journal={Physical Review Letters},
  volume={124},
  number={8},
  pages={080601},
  year={2020},
  publisher={APS}
}

@article{chubynsky2014,
  title={Diffusing diffusivity: a model for anomalous, yet Brownian, diffusion},
  author={Chubynsky, Mykyta V and Slater, Gary W},
  journal={Physical Review Letters},
  volume={113},
  number={9},
  pages={098302},
  year={2014},
  publisher={APS}
}

@article{benichou2011rev,
  title={Intermittent search strategies},
  author={B{\'e}nichou, O and Loverdo, C and Moreau, M and Voituriez, R},
  journal={Rev Mod Phys},
  volume={83},
  number={1},
  pages={81},
  year={2011},
  publisher={APS}
}

@article{vaccario2015,
  title={First-passage times in $d$-dimensional heterogeneous media},
  author={Vaccario, G and Antoine, C and Talbot, J},
  journal={Physical Review Letters},
  volume={115},
  number={24},
  pages={240601},
  year={2015},
  publisher={APS}
}

@article{grebenkov2016,
  title={Universal formula for the mean first passage time in planar domains},
  author={Grebenkov, Denis S},
  journal={Physical Review Letters},
  volume={117},
  number={26},
  pages={260201},
  year={2016},
  publisher={APS}
}

@book{redner2001,
  title={A guide to first-passage processes},
  author={Redner, Sidney},
  year={2001},
  publisher={Cambridge University Press}
}

@article{lawley2019dtmfpt,
title={Diffusive search for diffusing targets with fluctuating diffusivity and gating},
author={Lawley, S D and Miles, C E},
journal={Journal of Nonlinear Science},
year={2019},
  doi = {10.1007/s00332-019-09564-1},
  url = {https://doi.org/10.1007/s00332-019-09564-1},
  note={https://doi.org/10.1007/s00332-019-09564-1}
}

@article{bressloff2013stochastic,
	title = {Stochastic models of intracellular transport},
	volume = {85},
	url = {http://link.aps.org/doi/10.1103/RevModPhys.85.135},
	doi = {10.1103/RevModPhys.85.135},
	number = {1},
	urldate = {2016-06-16},
	journal = {Reviews of Modern Physics},
	author = {Bressloff, Paul C. and Newby, Jay M.},
	month = jan,
	year = {2013},
	pages = {135--196}
}

@incollection{chou2014first,
	title = {First passage problems in biology},
	booktitle = {First-{Passage} {Phenomena} and {Their} {Applications}},
	publisher = {World Scientific},
	author = {Chou, Tom and D'Orsogna, Maria R.},
	year = {2014},
	pages = {306--345}
}

@article{ward1993,
  title={Strong localized perturbations of eigenvalue problems},
  author={Ward, M J and Keller, J B},
  journal={SIAM J Appl Math},
  volume={53},
  number={3},
  pages={770--798},
  year={1993},
  publisher={SIAM}
}

@book{rayleigh1945,
  title={The theory of sound},
  author={Rayleigh, J W S},
  year={1945},
  publisher={Dover}
}

@article{berg1985orientation,
  title={Orientation constraints in diffusion-limited macromolecular association: {T}he role of surface diffusion as a rate-enhancing mechanism.},
  author={Berg, O},
  journal={Biophys J},
    volume={47},
  number={1},
  pages={1},
  year={1985},
  publisher={The Biophysical Society}
}

@article{berg1977physics,
  title={Physics of chemoreception},
  author={Berg, Howard C and Purcell, Edward M},
  journal={Biophys J},
  volume={20},
  number={2},
  pages={193--219},
  year={1977},
  publisher={Elsevier}
}

@article{shoup1982role,
  title={Role of diffusion in ligand binding to macromolecules and cell-bound receptors.},
  author={Shoup, David and Szabo, Attila},
  journal={Biophys J},
  volume={40},
  number={1},
  pages={33},
  year={1982},
  publisher={The Biophysical Society}
}

@article{collins1949diffusion,
	Author = {Collins, F. C. and G. E. Kimball},
	Journal = {J. Colloid. Sci.},
	Title = {Diffusion-controlled reaction rates},
	year = {1949},
	volume = {4},
	issue = {4},
	pages = {425--437}	
}

@article{holcman2014,
	title = {The Narrow Escape Problem},
	volume = {56},
	doi = {10.1137/120898395},
	pages = {213--257},
	number = {2},
	journal = {{SIAM} Rev},
	author = {Holcman, D and Schuss, Z},
	year = {2014}
}

@article{ward10b,
	title = {An asymptotic analysis of the mean first passage time for narrow escape problems: {P}art {II:} {T}he sphere},
	volume = {8},
	language = {en},
	number = {3},
	journal = {Multiscale Model Simul.},
	author = {Cheviakov, A. F. and Ward, M. J. and Straube, R.},
	year = {2010},
	pages = {836--870}
}

@article{ward10,
	title = {{An asymptotic analysis of the mean first passage time for narrow escape problems: Part I: Two-dimensional domains}},
	volume = {8},
	language = {en},
	number = {3},
	journal = {Multiscale Model Simul.},
	author = {Pillay, S. and Ward, M. J. and Peirce, A. and Kolokolnikov, T.},
	year = {2010},
	pages = {803--835}
}

@article{lawley2015new,
	title = {A new derivation of Robin boundary conditions through homogenization of a stochastically switching boundary},
	language = {en},
	journal = {SIAM J. Appl. Dyn. Syst.},
	volume = {14},
	number = {4},
	author = {Lawley, S. D. and Keener, J. P.},
	year = {2015}
}
\bibliographystyle{unsrt}

%%%%%%%%%%%%%%%%%%%%%%%%%%%%%%%%%%%%%%%%%%%%%%%%%%%%%%%%%%%%%%%%%%%%%%%%%%%%%%%%%%%%%%%%%%%%%%%%%%%%%%%%%%%%%%%%%%%%%%%%%%%%%%%%%%%%%%%%%%%%%%%%%%%%%%%%%%%%%%%%%%%%%%%%%%%%%%%%%%%%%%%%%%%%%%%%%%%%%%%%%%%%%%%%%%%%%%%%%%%%%%%%
\end{document}